\documentclass[structabstract]{aa}  
\usepackage{natbib}
\usepackage{graphicx}
\usepackage{txfonts}

\usepackage{amssymb}
\usepackage[notref,notcite]{} 

\newcommand{\T}{Tycho }
\newcommand{\cmt}[1]{}

\newcommand{\px}{p_{\max}}

\def\kms{~{\rm km~s^{-1}}}
\def\cm3{~{\rm cm^{-3}}}
\newcommand{\rxj}{RX J1713.7-3946}
\def\lesssim{\buildrel < \over {_{\sim}}}
\def\gtrsim{\buildrel > \over {_{\sim}}}

\begin{document}

\title{Strong evidences of hadron acceleration in Tycho's Supernova Remnant}

\author{G. Morlino \inst{1}\fnmsep\thanks{email: morlino@arcetri.astro.it}
        \and D. Caprioli\inst{1}\fnmsep\thanks{email: caprioli@arcetri.astro.it}
       }

\institute{INAF/Osservatorio Astrofisico di Arcetri, Largo E. Fermi, 5,
           50125, Firenze, Italy
          }

\date{Received August, 2011; accepted --, --}

 
  \abstract
   {Very recent gamma-ray observations of G120.1+1.4 (Tycho's) supernova remnant (SNR)
   by Fermi-LAT and VERITAS provided new fundamental pieces of information for understanding 
   particle acceleration and non-thermal emission in SNRs.}
   {We want to outline a coherent description of Tycho's properties 
   in terms of SNR evolution, shock hydrodynamics and multi-wavelength emission
   by accounting for particle acceleration at the forward shock 
   via first order Fermi mechanism.}
   {We adopt here a quick and reliable semi-analytical approach to non-linear diffusive shock acceleration
   which includes magnetic field amplification due to resonant streaming instability and the dynamical 
   backreaction on the shock of both cosmic rays (CRs) and self-generated magnetic turbulence.}
   {We find that Tycho's forward shock is accelerating protons up to at least 500 TeV, 
   channelling into CRs about the 10 per cent of its kinetic energy.
   Moreover, the CR-induced streaming instability is consistent 
   with all the observational evidences indicating a very efficient magnetic 
   field amplification (up to $\sim 300 \mu$G).
   In such a strong magnetic field the velocity of the Alfv\'en waves scattering CRs in
   the upstream is expected to be enhanced and to make accelerated particles feel an effective
   compression factor lower than 4, in turn leading to an energy spectrum steeper
   than the standard prediction $\propto E^{-2}$.
   This latter effect is crucial to explain the GeV-to-TeV gamma-ray spectrum
   as due to the decay of neutral pions produced in nuclear collisions between 
   accelerated nuclei and the background gas.}
   {The self-consistency of such an hadronic scenario, along with the fact that the 
   concurrent leptonic mechanism (inverse Compton scattering of relativistic electrons 
   on several photon backgrounds) cannot reproduce both the shape and the normalization 
   of the detected the gamma-ray emission, represents the first clear and direct 
   radiative evidence that hadron acceleration occurs efficiently in young
   Galactic SNRs.}

\keywords{acceleration of particles -- cosmic rays -- nonthermal emission --
          SNR:Tycho }

\maketitle

\section{Introduction}
The detection of high energy gamma-rays from supernova remnants (SNRs) has long
been considered as the most promising tool \cite[see e.g.][]{dav94} to probe the
so-called \emph{supernova paradigm} for the origin of Galactic cosmic rays
(CRs), which states that SNRs are responsible for the acceleration of nuclei up
to energies as high as a few times 10$^6$ GeV \citep[see e.g.][for a
review]{Hillas05}. 

In the last few years several SNRs have been detected in gamma-rays, both in the
GeV and in the TeV band (see \cite{Caprioli11} for a recent summary) but, in
spite of this larger and larger amount of data, a clear-cut evidence that such
an emission is due to the decay of neutral pions produced in nuclear
interactions between accelerated nuclei and the background plasma is still
lacking. The very reason why the probe of such a \emph{hadronic scenario} for
the origin of the detected gamma rays is so controversial is that also processes
involving relativistic electrons (\emph{leptonic scenario}) may provide a
similar gamma-ray signature \citep[see e.g.][for a general discussion on this
topic]{Ellison07}.

The best example of this issue is indeed represented by the SNR RX J1713.7-3946,
which for many years has been regarded as the best candidate as a hadron
accelerator due to its brightness in the TeV band \citep[see e.g.][]{Morlino09,
ZirAha10}. 
Nevertheless, very recent high-resolution observations in the GeV band performed
by Fermi-LAT \citep{Abdo11} showed that the emission from the SNR shell is more
likely due to Inverse-Compton scattering (ICS) of relativistic electrons rather
than to pion decay. Its spectrum in the GeV band is in fact very flat ($\propto
E^{-1.5}$), as predicted by a "leptonic" model, which is also supported by 
the lack of detection of thermal X-ray emission \cite{Ellison10-RXJ}.

Another leptonic mechanism which may be relevant for the gamma-ray emission from 
SNRs is the relativistic bremsstrahlung of accelerated electrons which,
always coupled with ICS, has been proposed as a viable alternative to the
hadronic scenario to explain for instance the spectrum of Cas A \citep{Abdo10}.
There are several other examples of SNRs whose emission could be, and has been, 
accounted for either in an hadronic or in a leptonic scenario \citep[see][and
references therein]{Caprioli11} but, at the moment, for no gamma-ray-bright
source the leptonic scenario has been ruled out beyond a reasonable doubt. 

In several cases the extended TeV emission has been suggested to originate from
the interaction between particles accelerated in SNR and nearby molecular
clouds (MCs), which naturally provide a large number of target nucleons. 
This scenario may be preferentially associated with remnants produced by
core-collapse SNe, whose progenitors evolve very quickly in star formation
regions still populated with parent MCs. In these environments, however, it is
not very clear whether to expect large CR acceleration efficiencies, because of
the small ionized fraction and, likely, because of ion-neutral damping
\cite[]{Drury96}.

As pointed out for instance by \cite{Caprioli11}, the best sources where to look
for hadron acceleration signatures are gamma-ray-bright SNRs not associated with MCs and
showing rather steep spectral indexes both in the GeV and in the TeV bands.
We identify Tycho's SNR, freshly detected by Fermi-LAT and VERITAS, as the best
candidate for investigating the effects that particle acceleration
produces at all observable wavelengths, simultaneously using multiple set of data
to constraint our model. 
To discuss things consistently, we nevertheless need to adopt a very
refined model able to account, at the same time, for the SNR evolution, the
acceleration of particles at the shock and, as it will be clearer soon, the magnetic
field amplification produced by the streaming of the same relativistic particles. 

In this work we therefore apply the non-linear diffusive shock acceleration
(NLDSA) theory to the Tycho's SNR following the semi-analytical approach put 
forward by \cite{ab05,ab06,dsax0,crspec} and references therein. Thanks to the
large amount of data available at various wave bands, this remnant can be
considered one of the most promising object where to test the shock acceleration
theory and hence the CR--SNR connection. 
Tycho is among the youngest Galactic SNRs, being only 439 years old: recent
observations have in fact confirmed that it is the remnant of a Type Ia SN
exploded in 1572 by detecting its optical spectrum near maximum brightness from
the scattered-light echo \cite[]{Krause08}. 

\begin{figure}
\begin{center}
{\includegraphics[width=.8\linewidth]{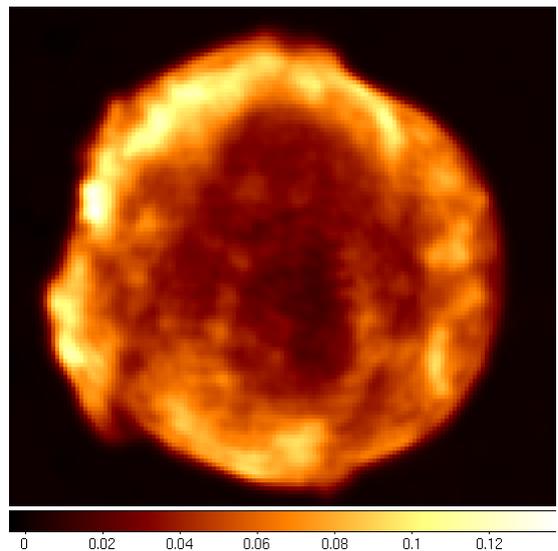}}
  \caption{Radio image of the Tycho's remnant at 1.5 GHz in linear color scale.
  Image credit: NRAO/VLA Archive Survey, (c) 2005-2007 AUI/NRAO.}
  \label{fig:Radio_map}
\end{center}
\end{figure}

Tycho shows a radio spectral index of 0.6-0.65 and a flux density as large as 40.5 Jy
at 1.4 GHz \cite[]{Kothes06}. 
Very interestingly, such a spectral index requires electrons in the GeV energy
range to have a spectral slope $\sim 2.2-2.3$, i.e.\ steeper than the standard
prediction of linear acceleration theory. The radio morphology is clearly
shell-like, with enhanced emission along the northeastern edge (see
Fig.~\ref{fig:Radio_map}). A hint of curvature in the radio spectrum, consistent
with non-linear theories' expectations, has also been reported by
\cite{ReyEllison92}, who also inferred a magnetic field strength $\gtrsim 100
\mu$G.

In gamma-rays, instead, Tycho has been observed several times with no credible
detection reported until only very recently the VERITAS Cherenkov telescope 
succeeded in measuring its TeV emission up to 10 TeV \cite[]{Acciari11}, 
providing an integral flux over 1 TeV of only $\sim 0.9\%$ of the steady Crab
Nebula emission above the same energy. 
Such a low emission makes Tycho one of the weakest TeV source ever detected.
The TeV emission is compatible with a point like source since the telescope's 
point spread function (PSF) of $\sim 15$ arcmin is larger than the angular size
of the remnant (8 arcmin). Even more recently, the {\it Fermi} Large Area
Telescope (Fermi-LAT) has reported a detection of Tycho as well, assessing an
integral flux above 400 MeV of $\sim (3.5\pm 1.1_{stat}\pm0.7_{sys}) \times
10^{-9} \rm ph\, cm^{-2} s^{-1}$ and a spectral slope
$2.3\pm 0.2_{stat} \pm 0.1_{sys}$\cite[]{Giordano11}.

It has been raised the question whether the gamma-ray emission were correlated
with the interaction of the remnant with a nearby MC.  The peak of the TeV
emission might in fact be slightly displaced northeast to the remnant's center,
where also a CO cloud has been detected, but the statistical significance of
this displacement seems rather low \cite[]{Acciari11}. A possible interaction
between the northeastern part of Tycho and dense molecular gas has
indeed been suggested by many authors \cite[]{Reynoso99,Lee04,Cai09,Xu11}, but
in a more recent study \cite{Tian11} analyzed the 21 cm continuum, the HI and
$^{12}$CO-line data, concluding that there are no evidences of a direct
interaction between the shock and a MC.

Interestingly enough, \T is also clearly visible in the infra-red (IR) band as
well \cite[]{Ishihara10}. In particular, mid-IR images show a shell-like
structure with brightness peaks at the northeast and northwest boundaries,
probably due to dust collisionally heated up by the hot shocked plasma in the
SNR downstream. The far IR band, instead, is dominated by cold dust emission, 
which spatially correlates with the CO MC mentioned above, hence it is probably
not related to the remnant. IR emission due to heated dust is particularly
important because it provides the principal photons background for the ICS
of accelerated electrons, as we will show in \S\ref{sec:gamma}.

High-resolution X-ray maps reveal strong non-thermal emission concentrated in
thin filaments \cite[]{Hwang02,Bamba05,Warren05,Katsuda10,Eriksen11}. 
The thickness of these filaments is usually interpreted as due to severe
synchrotron losses of high energy electrons radiating in a strong magnetic field
\cite[e.g.][]{Ballet06}. 
This interpretation allows us to estimate the magnetic field strength, which has
to be as large as $\sim 250 \mu$G inside the rim \cite[]{Volk05,Parizot06}.
There is however also the possibility that the observed X-ray morphology may
result from a rapid damping of the magnetic field in the downstream
\cite[]{Pohl05}. \cite{MarcoCasse10} investigated this scenario considering
different magnetic relaxation processes and different kind of magnetic
turbulence and showed that, in order to have a relaxation length comparable to
the X-ray rim thickness, the downstream magnetic field has invariably to be 
of order $200-300\mu$G. 
Smaller fields would in fact require much longer distance to be damped. Hence,
regardless of the presence of any damping mechanisms, the small thickness of
X-ray filaments can be regarded as a proof that the magnetic field is amplified
from the typical interstellar value up to hundreds of $\mu$G. 
In \S\ref{sec:damp} we show that the presence of a damping mechanism for the
magnetic field can be investigated also by analyzing the radial
profile of the radio emission.

It is interesting to recall that detailed X-ray morphology studies have already
provided evidences of efficient hadron acceleration in Tycho. In fact
\cite{Warren05} estimated the location of the contact discontinuity (CD) and the
forward shock (FS) by comparing thermal and non-thermal X-ray emissions, 
concluding that they are too close to be described by purely gaseous 
hydrodynamical models. They therefore argued that the compression factor at the
FS had to be  larger than 4, as predicted in the case of efficient CR
acceleration.

Finally we note that hints on the presence of accelerated ions come also from
the detection of H$\alpha$ lines. Analyzing some Balmer dominated filaments in
the eastern limb of the remnant, \cite{Lee10} found a significant
amount of the H$\alpha$ emission to come from the region upstream of the shock,
which has been interpreted as a CR-induced precursor.

All these things considered, several observations have already provided strong hints 
that efficient CR acceleration is occurring at Tycho's FS. 
Our goal is therefore to provide a self-consistent picture able
to take into account all of these observational evidences and to couple them 
with brand new gamma-ray data. 
The NLDSA theory has already been adopted in order to predict the multi-wavelength
spectrum of Tycho \cite[e.g.][]{Volk05,Volk08}, nevertheless here we improve on
previous attempts in some important aspects: 
1) the magnetic field upstream of the shock is calculated from resonant streaming instability; 
2) the dynamical reaction of the magnetic field onto the shock dynamics is accounted for; 
3) the upstream scattering center velocity is properly calculated in the amplified magnetic field; 
4) the non-linear Landau damping of the magnetic field in the downstream is introduced; 
5) the ICS of accelerated electrons is calculated considering as target photons non only the 
cosmic microwave background (CMB) radiation, but also the Galactic background and, more importantly, the IR photons produced by the local warm dust.

The inclusion of the dynamical reaction of the field reduces the compressibility
of the plasma and affects the prediction for the shock compression factor
\cite[]{jumpkin}. 
A crucial ingredient is the velocity of the scattering
centers, which is generally neglected with respect to the shock speed, but could
be significantly enhanced when the magnetic field is amplified \cite[]{veb06,jumpkin,zp08}. 
When this occurs, the total compression factor felt by accelerated particles may be
appreciably reduced and, in turn, the spectra of accelerated particles may be considerably softer.

It is worth remembering that some observational features, especially the radio
emission, are strongly affected by the past history of the remnant, hence any
reliable calculation has to take into account also the SNR evolution. 
In this paper we use a stationary version of NLDSA theory, but we couple this theory
to the hydrodynamical evolution of the remnant provided by \cite{TMK99}. 
We divide the SNR evolution in several time steps and we assume that for each time step the
stationary theory can be applied, like has been done in \cite{crspec}.
However, as showed by \cite{comparison}, stationary models and time-dependent
approaches return very similar CR spectra for non-relativistic shocks.

We compare the results of our kinetic model with the multi-wavelength 
integrated spectrum of Tycho from the radio to the TeV range, 
and also with the radial profile of X-ray and radio emissions.
Our conclusion is that existing data of Tycho's SNR are consistent with a
moderately efficient acceleration of CR nuclei: 
at the present age we infer that a fraction around 12 per cent of the total 
kinetic energy has been converted in CRs. 
Such an efficiency also implies an amplified magnetic field of 
$\sim 300\mu$G, perfectly consistent with the measured X-ray rim thickness.
In addition, such a strong magnetic field enhances the velocity of 
the scattering centers, finally reducing the effective compression factor 
felt by accelerated particles, whose spectrum turns out to be as steep as  
$\sim E^{-2.2}$. 
The most important consequence of this fact is that this spectrum allows us to fit
the observed gamma-ray emission, from the GeV to the TeV band, as due to neutral pion
decay.  
Moreover, in this framework it is not possible to explain the TeV emission as due to ICS
without violating many other observational constraints.

The paper is organized as follows: in \S\ref{sec:model} we summarize the
details of our model for non-linear particle acceleration and
our treatment of the SNR evolution. 
In \S\ref{sec:results} we outline the macroscopic properties of Tycho's SNR, in
order to fix the free parameters of our model, while
in \S\ref{sec:spectrum} we widely discuss the comparison between data and our 
findings for the multi-wavelength spectrum, also by analyzing each different
energy band separately. We conclude in \S\ref{sec:conclusions}.

\section{Description of the model}
\label{sec:model}

\subsection{Remnant evolution}\label{sec:hydro}
We model the evolution of Tycho by following the analytic prescriptions given by
\cite{TMK99}. More precisely, we consider a SN explosion energy
$E_{SN}=10^{51}$ erg and one solar mass in the ejecta, whose structure function is taken as
 $\propto (v/v_{ej})^{-7}$ \citep[see \S3.2 and \S9 in][]{TMK99}. 
Such a set of parameters has been showed to be suitable for describing the
evolution of the FS position and velocity for a type Ia SNR: the
parametrization given in table 7 of \cite{TMK99} in fact differs
from the exact numerical solution of about 3 per cent typically, 
and of 7 per cent at most.
Such a solution, which does not include explicitly the possible role of
the CR pressure in the SNR evolution, is still expect to hold for moderately
small acceleration efficiencies (below about 10 per cent).
We checked a posteriori that the efficiency needed to fit observations does not
require amore complex treatment of the shock evolution during the
ejecta-dominated stage.

\begin{figure}
\begin{center}
{\includegraphics[width=\linewidth]{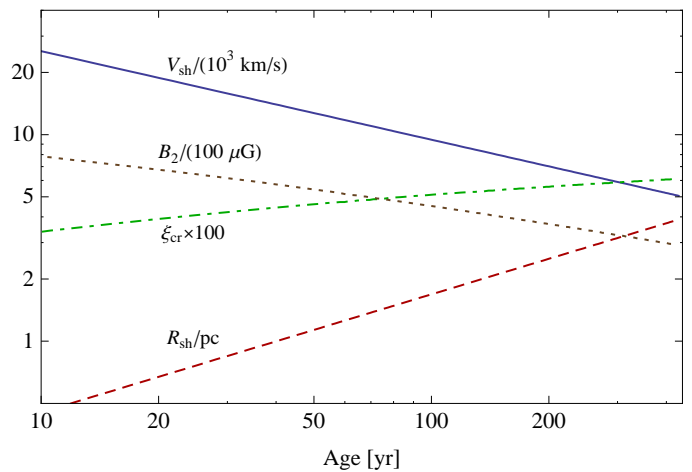}}
\caption{Time evolution of shock radius $R_{sh}$, shock velocity $V_{sh}$, magnetic field
immediately behind the shock $B_2$ and CR acceleration efficiency $\xi_{cr}=P_{cr}/\rho_0V_{sh}^2$.}
\label{fig:evo}
\end{center}
\end{figure}

The circumstellar medium is taken as homogeneous with proton number density 
$n_{0}=0.3$ cm$^{-3}$ and temperature $T_{0}=10^{4}~{\rm K}$. 
Following the conclusion of \cite{Tian11}, we assume that the remnant
expands into the uniform interstellar medium (ISM) without interacting with any MC.
With these parameters, the reference value for the beginning of the
Sedov-Taylor stage is $T_{ST}=463$yr, therefore \T is experiencing the
transition between the ejecta-dominated and the adiabatic stage. 
FS radius and velocity can be hence simply written as \citep{TMK99}:
\begin{eqnarray}
&R_{sh}(t)& =4.06\left(\frac{t}{T_{ST}}\right)^{4/7} {\rm pc} \,,\label{tmr} \\
&V_{sh}(t)& =4875\left(\frac{t}{T_{ST}}\right)^{-3/7} {\rm km~s^{-1}}\label{tmv}\,.
\end{eqnarray}

The time evolution of some relevant physical quantities is depicted in
Fig.~\ref{fig:evo}. The last point corresponds to an age of 439 yr and hence to
our estimate for the present shock velocity, $V_{sh}\simeq 4990\kms$, and shock
radius, $R_{sh}=3.94$ pc.

The radial structure of density and temperature profiles is then calculated
by assuming that the downstream (from the CD to the FS), being subsonic, is
roughly in pressure equilibrium. In our case, this recipe leads to a
discrepancy of less than 10 per cent with respect to exact profile calculated by
including also the contribution of the CR pressure in the SNR evolution
\citep[see e.g.][]{chevalier83}.

More precisely, since $\rho^{\gamma}(t)\propto p_{gas}(t)\propto V_{sh}^{2}(t)$,
the adiabatic decompression of a fluid element with density $\rho_{0}$ shocked
at time $t_{0}$ can be calculated as 
\begin{equation}\label{eq:L}
\frac{\rho(t)}{\rho_{0}}=L(t_{0},t)^{3}=
  \left[ \frac{V_{sh}(t)}{V_{sh}(t_{0})} \right]^{\frac{2}{\gamma}},
\end{equation}
where $\gamma$ is the effective adiabatic index of the CR+plasma fluid. 
In general we have $4/3\leq \gamma \leq 5/3$, but in our case $\gamma=5/3$ is a very
reasonable choice, justified by the fact that the CR acceleration efficiency we
find is never much larger than $\sim 10$ per cent. 
The quantity $L(t_{0},t)\leq 1$ introduced above represents the 
adiabatic energy loss CRs and magnetic fields
undergo in the period $t-t_{0}$, and it is consistently taken into account in
modelling the emission from fluid elements shocked at earlier times.

\begin{figure}
\begin{center}
{\includegraphics[width=\linewidth]{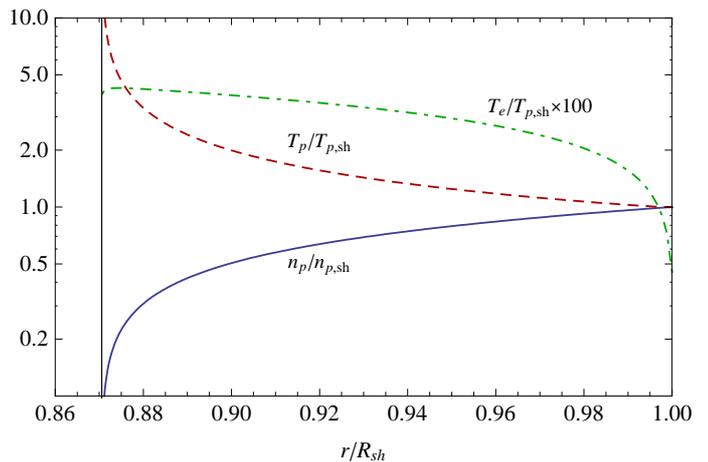}}
\caption{Density (solid line) and temperature (dashed line) of the shocked
ISM in the region between the CD (vertical solid line) and the FS, computed at the
present age of the remnant, $t=439$ yr, and normalized to the values
immediately behind the shock. The dot-dashed line shows the electron temperature
computed including the heating due to Coulomb collisions.}
\label{fig:hydrofin}
\end{center}
\end{figure}

In a similar way, this simple recipe allows us to track also the downstream
temperature profile and therefore to work out the bremsstrahlung emission due to
thermal electrons, which in general have a temperature $T_e$ different from
the protons' one, $T_p$. We assume $T_{e}= T_{p}\,m_e/m_p$ immediately after the
shock and calculate the heating of electrons downstream as due to Coulomb collisions
against hotter protons. 
In Fig.~\ref{fig:hydrofin} we plot the value of
electron and proton temperatures between the CD and the FS at the present epoch,
normalized to the proton temperature immediately behind the shock,
$T_{p,{\rm sh}}= 5\times10^8$ K. 
The figure shows that Coulomb collisions rapidly enhance the electron temperature 
up to few per cent of $T_{p,{\rm sh}}$, and also that the maximum
value is reached close to the CD, where we have $T_{e}\simeq 2\times 10^7$ K.
We would like to stress that Coulomb collisions provide the minimum heating rate for
electrons; others plasma processes could in principle equilibrate protons and
electrons even more rapidly. On the other hand, as we will see in
\S\ref{sec:X-rays}, the thermal electron bremsstrahlung predicted by assuming
Coulomb heating only is compatible with the non detection of X-ray thermal emission
\citep{Gamil07}. 

\subsection{Particle acceleration}\label{sec:acceleration}
On top of this SNR evolution, the spectrum of accelerated particles is calculated
according to the semi-analytic kinetic formalism put forward in \cite{dsax0}
and references therein (especially \cite{ab05,ab06}), which solves
self-consistently the equations for conservation of mass, momentum and energy
along with the diffusion-convection equation describing the transport of
non-thermal particles for quasi-parallel, non-relativistic shocks.
In particular, we impose the CR distribution function to vanish at a distance
$\sim \chi_{esc} R_{sh}$ upstream of the shock, mimicking the presence of a
free-escape boundary beyond which highest-energy particles cannot diffuse back
at the shock and get lost in the ISM \cite[]{dsax0}.
This constraint actually determines the instantaneous maximum momentum  
than can be achieved by CRs by imposing the diffusion length 
of particles with momentum $p_{\max}(t)$ to be comparable with the distance
between the shock and the free escape boundary, namely
\begin{equation}\label{eq:pmax}
\frac{D(p_{\max})}{V_{sh}}= \chi_{esc}R_{sh}\simeq 0.1 R_{sh}\,,
\end{equation}
where $D(p)$ is the diffusion coefficient, here taken as Bohm-like, i.e.\ $D(p)=
\frac{v(p)}{3}r_{L}(p)$, with $v(p)$ and $r_{L}(p)$ the particle velocity and the Larmor
radius, respectively.

We also checked a posteriori that by posing $\chi_{esc}=0.1$ such a
size-limited $p_{\max}$ is also consistent with requiring both the acceleration time
up to $p_{\max}$ to be smaller than the age of the system \cite[]{maximum}
and the diffusion length downstream to be smaller than the distance between FS and CD.

Particles are injected into the acceleration mechanism from the suprathermal
tail of the Maxwellian distribution (thermal leakage) following the prescription by \cite{bgv05}. 
We assume that all the particles with a momentum larger than
$\xi_{\rm inj}$ times the downstream thermal momentum have a Larmor radius large enough
to cross the shock from downstream to upstream, and in turn to be injected in
the acceleration process.
We fix $\xi_{\rm inj}=3.7$, corresponding to let a fraction $\sim 10^{-4}$ of the particles
crossing the shock to be accelerated as CRs.

\begin{figure}
\begin{center}
\includegraphics[width=\linewidth]{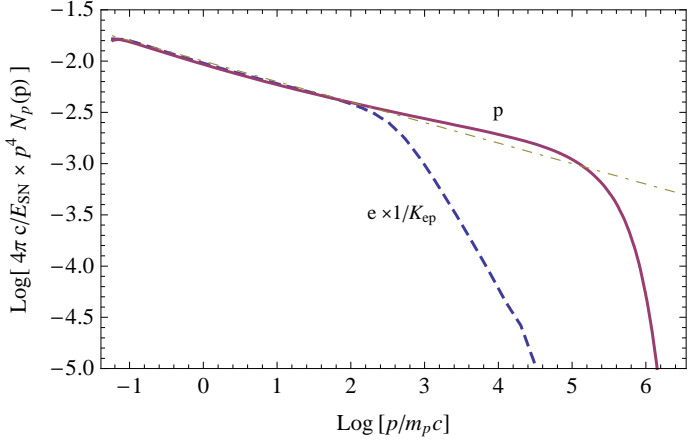}
\caption{Total spectrum of protons and electrons (divided by $K_{ep}$) at the
present age of the remnant, as a function of the particle momentum.  For
comparison, the dot-dashed line shows a power-law $\propto p^{-4.2}$ which
emphasizes a small flattening of the proton spectrum at the highest momenta.} 
\label{fig:p-e}
\end{center}
\end{figure}

It is worth stressing that the solution of the NLDSA problem obtained with the
semi-analytical approximate solution worked out in \cite{dsax0} is in excellent
agreement with the exact solution of the stationary diffusion-convection
equation for CR transport coupled with conservation equations, but it is also
very consistent with state-of-the-art numerical approaches to the time-dependent
problem and also with Monte Carlo approaches able to retain all the information
about the anisotropy of the whole (thermal particles + CRs) distribution
function \citep{comparison}.

\subsection{Magnetic field amplification} \label{sec:field}
The super-Alfv\'enic streaming of relativistic particles is expected to excite a
plasma instability (\emph{streaming instability}) which strongly enhances
Alfv\'en waves resonant with the Larmor radii of relativistic particles
\citep{skillinga,bell78}. Such a magnetic turbulence may grow well beyond the
quasi-linear limit, eventually producing a perturbation $\delta B\gg B_{0}$,
where $B_{0}$ is the component of the ordered background magnetic field parallel
to the CR gradient, and hence to the shock normal.
This process, usually referred to as magnetic field amplification, has rather
solid observational counterparts in the X-ray observations of many
young SNRs: the downstream narrow X-ray rims due to severe synchrotron losses by
relativistic electrons \citep[see e.g.][]{Volk05,Parizot06}, the rapid
variability of the X-ray-bright spots in \rxj~\citep{Uchiyama07} and also the
limited extension of the upstream emission in SN~1006 \citep{mor10}.

\T makes no exception, showing evidences of narrow non-thermal X-ray rims
pointing to magnetic fields as large as a few hundreds $\mu$G
\citep{Hwang02,Bamba05,Warren05,Katsuda10,Eriksen11}. 

We model magnetic field
amplification due to resonant streaming instability as in \cite{jumpkin}, i.e., by
assuming that saturation is achieved when $P_{w}\simeq P_{cr}/2M_{A}$, where
$P_{w}$ and $P_{cr}$ are the pressure in Alfv\'en waves and in CRs, normalized
to the ram pressure $\rho_{0}V_{sh}^{2}$, and $M_{A}$ is the Alfv\'enic Mach
number $M_{A}=V_{sh}/v_{A}=V_{sh}\sqrt{4\pi\rho_{0}}/B$.
In the limit $M_{0}^2\gg1, M_{A}^2\gg1$ the solution for the
wave transport equation reads \citep[see eq.~43 in][]{jumpkin}: 
\begin{equation}\label{saturationSI}
 P_{w}(x)=
  \frac{B(x)^{2}}{8\pi\rho_{0}V_{sh}^{2}}\simeq
  \frac{1+U(x)}{4M_{A}(x)U(x)} \, P_{cr}(x),
\end{equation}
where $U(x)$ is the fluid velocity (normalized to $V_{sh}$) in the upstream precursor.

When magnetic field amplification occurs a strong non-linear regime, as it is
expected in our case, it is not clear whether the magnetic field entering the
Alfv\'en velocity has to be the background one, $B_{0}$, as predicted by an
extrapolation of the quasi-linear theory, or the amplified one, $\delta B\gg
B_{0}$. Strictly speaking, the quasi-linear prediction is related to the fact
that resonant transverse Alfv\'en waves ($\delta \vec{B}\perp \vec{B}_{0}$) of
arbitrary strength are exact solutions of the problem: it is however very likely
that some mechanism may tend to make the field rather turbulent when $\delta
B\gg B_{0}$, therefore we choose to calculate the Alfv\'en velocity in the
amplified magnetic field.
The implications of this choice are discussed more widely in \cite{Caprioli11}.

Here and in the following we label with the subscript 1 (2) quantities measured
immediately ahead (behind) the shock, while the subscript 0 is reserved for
undisturbed quantities at upstream infinity. The maximum value of the upstream
magnetic field is thus reached immediately ahead of the shock and it is given
by 
\begin{equation}
 P_{w,1}=
  \frac{B_{1}^{2}}{8\pi\rho_{0}V_{sh}^{2}}= 
  \frac{1+U_{1}}{4M_{A,1}U_{1}}\, \xi_{cr},
\end{equation}
where we posed $\xi_{cr}=P_{cr,1}\simeq 1-U_{1}$. The relevant Alfv\'enic Mach
number thus reads 
\begin{equation}\label{eq:Ma1}
 M_{A,1}\simeq\frac{2U_{1}^{2}}{1-U_{1}^{2}}=\frac{2}{\xi_{cr}}\frac{(1-\xi_{cr})}{2-\xi_{cr}}^{2}
\end{equation}
and therefore we have upstream of the subshock:
\begin{equation}\label{B1}
 B_{1}= \frac{\sqrt{4\pi\rho_{1}}U_{1}V_{sh}}{M_{A,1}}=
   \sqrt{\pi\rho_{0}} \, V_{sh} \,
   \frac{\xi_{cr}(2-\xi_{cr})}{(1-\xi_{cr})^{3/2}} \,,
\end{equation}
where we also used the mass conservation $\rho_{1}U_{1}=\rho_{0}$.

We then assume the downstream magnetic field to have a strength $B_{2}=
\sqrt{(1+2R_{sub}^{2})/3}\, B_{1} \simeq \sqrt{11}\, B_{1}$, where $R_{sub}
\simeq 4$ is the subshock compression ratio, since only the components
perpendicular to the shock normal are compressed. For the shock parameters
considered above ($\rho_{0}=0.3m_{p}$cm$^{-3}$, $V_{sh}\simeq 5000\kms$) a CR
acceleration efficiency of about 5--10 per cent is therefore enough to provide a
downstream magnetic field of a few hundreds $\mu$G (see Eq.~\ref{B1}), in
agreement with the values inferred by X-ray observations of young remnants.

As outlined in the introduction, such large magnetic fields have additional
consequences on both the shock hydrodynamics and on the spectrum of accelerated
particles. More precisely, when the upstream pressure in magnetic fields becomes
comparable to, or even larger than, the thermal plasma pressure, the global
dynamics of the system is regulated by the interplay between the CR  pressure
and the pressure in magnetic turbulence \citep{jumpl, jumpkin}.
To be more quantitative, at the shock $P_{w,1}\simeq 1/2M_{A,1}^{2} \approx
2.3\times 10^{-3}$ for $\xi_{cr}=0.06$ and hence $M_{A,1}\simeq 15$, which is
much larger than $P_{g,1}\simeq 1/\gamma M_{0}^{2}\approx5.6\times 10^{-6}$,
since the sonic Mach number is $M_{0}\simeq443$.
The additional pressure and energy density in magnetic field thus help to
balance the compression of the upstream fluid induced by CR pressure and is
consistently included both in the conservation equations and in the calculation
of the jump conditions at the subshock, as described by \cite{jumpl,jumpkin}.
Moreover, such a magnetic feedback helps in keeping the compression
ratios rather close to 4, which in turn legitimates the adoption of
Eqs.~\ref{tmr} and \ref{tmv}.

The second, and probably most important, effect of the efficient magnetic field
amplification is that the phase velocity of the Alfv\'en waves CRs scatter against 
may become a non-negligible fraction of the fluid velocity. In particular, Alfv\'en
waves produced by the CR gradient via streaming instability travel in opposite
direction with respect to the fluid, so that upstream CRs actually feel a net
velocity $\tilde{u}=u-v_{A}=u(1-1/M_{A})$, while downstream it is likely for
helicity mixing to lead to $v_{A,2}\approx 0$.
Since for DSA the spectral slope is determined only by the compression ratio
$\tilde{r}\simeq (u_1-v_{A,1})/u_2=u_1/u_2(1-1/M_{A})$ felt by diffusing particles, 
the non-negligible velocity of the scattering
centers (often also called \emph{Alfv\'enic drift}) naturally leads to particle
spectra steeper than what predicted without including this effect.

The potential role of Alfv\'enic drift in the DSA theory has already been
pointed out by \cite{bell78}, but it is important to stress that steep particle
spectra can be produced at SNR shocks only if the magnetic field is amplified 
via some instability to levels corresponding to $M_{A}\simeq 20$ or lower
\citep[see also][]{Caprioli11}. In our case, taking the saturation of the
streaming instability in the amplified field makes this phenomenon even more
non-linear, with the final result that the larger the number of particles
injected, the stronger the amplified magnetic field and eventually the steeper
the CR spectra achieved in a consistent NLDSA calculation \citep{Texas10}. 

Finally, another fundamental aspect of magnetic field amplification is that
diffusion in the enhanced magnetic turbulence allows particles to
return to the shock more quickly, in turn achieving a much larger momentum both
in the age-limited and in the space-limited scenario (see \S\ref{sec:acceleration}).
In our calculations we therefore take the Bohm-like diffusion coefficient in the
amplified magnetic field $B_{1}$ rather than in $B_{0}$, finding
$p_{\max}$ as high as $\sim 500$ TeV for the accelerated protons.  

Nevertheless, the excitation of modes resonant with CR gyroradii is not the only 
phenomenon which may be responsible for magnetic field amplification.
As shown by \cite{bell04} and then by \cite{mario-anatoly10}, also non-resonant modes
with short wavelengths may effectively grow because of CR escaping the system.
The relative importance of resonant and non-resonant modes has been investigated
within a kinetic approach by \cite{ab09}, who found that for young, strong
shocks also non-resonant modes may play a non-negligible role.
A comprehensive treatment of the interplay between CRs and magnetic fields from a kinetic point of
view is however beyond the goal of this paper, therefore we simply assume a uniform
magnetic field in the upstream given by Eq.~(\ref{B1}).
Such a recipe is indeed very reasonable in the shock proximity and becomes more
and more heuristic in the far upstream, where amplification should be
prevalently due  to the excitation of Bell's modes. 
However, if far upstream the magnetic field were much lower than we assumed, we
would have two effects: on one hand, the CR spectrum would be slightly flatter
($\propto E^{-1.95}$) at the highest energies, where particles would feel a
total compression ratio of about $R_{tot}\simeq 4.2$ and, on the other hand, the
maximum energy would be consistently reduced. 
These two effects would indeed have a radiative signature in the TeV band but
the quite large measurement errors and the lack 
of detection of a high-energy cut-off do not allow us to make a
realistic investigation of a possible flattening at the highest energies, yet.

In addition, also long-wavelength modes may play a role in the problem enhancing
the diffusion of the highest-energy particles \citep[see, e.g.,][and references
therein]{b11} up to PeV fractions. Our simple choice, very common in the
literature, of adopting a constant magnetic field upstream is indeed an
oversimplification of the problem but, within our incomplete understanding of
the mechanisms responsible for the scattering of the highest-energy particles,
allows us to heuristically account for a crucial observational evidence: if the
GeV-to-TeV emission is hadronic, as we will demonstrate in the following,
protons have to be accelerated up to at least a few hundreds TeV \citep[see
also][]{Eriksen11}.

\subsection{Magnetic field damping}\label{sec:damp}
As it will be clearer in the following discussion, in order to reproduce both
the intensity and the radial profile of the synchrotron emission, it is
necessary to account for some magnetic field damping in the downstream as well.
Several damping mechanisms have been proposed to be effective in SNRs \citep[see
e.g.][and references therein]{pz03}, but here we focus only on non-linear Landau
damping, which is expected to be rather efficient in hot (low Mach number)
plasmas, as outlined by \cite{v-mck81,mck-v82}.

We therefore assume that in the downstream the magnetic field is
damped with a rate, averaged on the different scales, given by \citep[see
eqs.~10--12 in][]{pz03}:
\begin{equation}
 \Gamma_{nl}\simeq \, 0.05 \, k_{\min} v_{A} \simeq
0.05 \frac{v_A}{r_L(p_{max})} = 
  \frac{0.05}{3}\frac{v_{A}(B_1)c}{\chi_{esc}V_{sh}R_{sh}}\,,
\end{equation}
where we took the maximum turbulence scale (i.e.\ the minimum wave number
$k_{\min}$) to be given by the Larmor radius of particles with momentum close to
$\px$ (in the last equality we simply made use of Eq.~\ref{eq:pmax}).

Since $B_1U_1\simeq B_2 U_2$, the typical length-scale for the non-linear
Landau damping is thus given by
\begin{equation}
 \lambda_{nl}\simeq \frac{U_2V_{sh}}{\Gamma_{nl}}=
   \frac{3\chi_{esc}}{0.05}\frac{V_{sh}^{2}}{v_{A}(B_{2} )c}R_{sh}
   \approx 3.5\, {\rm pc} \,,
\end{equation}
and therefore we obtain the following recipe for the downstream magnetic field: 
\begin{equation} \label{eq:B_damp}
 B(r)\simeq B_{2}\exp{ \left( -\frac{R_{sh}-r}{\lambda_{nl}}\right)}\,.
\end{equation}

\begin{figure}
\begin{center}
\includegraphics[width=\linewidth]{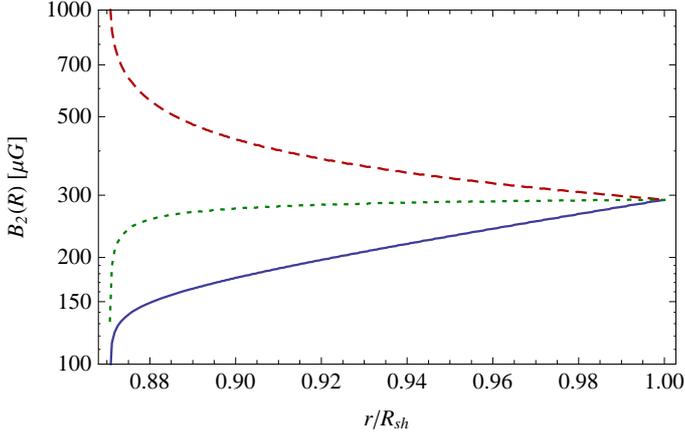}
\caption{Radial profile of the magnetic field in the downstream at the present
age of the remnant. The dashed curve shows $B_2$ as result from the
amplification and compression at the shock. 
The dotted curve include adiabatic losses, 
while the solid curve is computed including both the adiabatic
expansion and the Landau damping.} 
\label{fig:Bfield}
\end{center}
\end{figure}

The actual radial profile of $B(r)$ is finally showed in Fig.~\ref{fig:Bfield}
(solid line) and compared with the undamped magnetic field (dashed line) and
with the magnetic field subjected only to adiabatic expansion (dotted line).
 
\subsection{Electron spectrum} \label{sec:electron}
Once the proton spectrum has been calculated self-consistently with the shock 
dynamics, it is also possible to work out the spectrum of accelerated electrons,
which is parallel to the protons' one since DSA is charge independent, but
truncated at a lower maximum energy determined by synchrotron losses in the
amplified magnetic field.

The instantaneous electron spectrum at the shock, $f_{e,0}(p)$, is computed as
in \cite{Morlino09}, using the expression worked out by \cite{ZirAha07} for a
$\propto p^{-4}$ proton spectrum:
\begin{equation} \label{eq:f_e(p)}
 f_{e,0}(p)= K_{ep}\,f_{p,0}(p)
  {\left[1+0.523 \left(p/p_{e,\rm max}\right)^{\frac{9}{4}}\right]}^2
  e^{-p^2/p_{e,\rm max}^2} \,,
\end{equation}
where $p_{e,\max}$ is determined by equating the acceleration time with the
energy loss timescale due to synchrotron emission, which, as a consequence of
the efficient magnetic field amplification, turns out to be much smaller than
Tycho's age for large momenta \citep[see Eqs.~(3) and (4) in][]{Morlino09}.
Such a spectrum is expected to be a reasonable approximation of the true one 
\citep[see e.g.][]{blasi10}, since our proton spectrum is almost a power-law only 
slightly steeper than $p^{-4}$, as showed in Fig.~\ref{fig:p-e}.

A very important property of Eq.~(\ref{eq:f_e(p)}) is also to retain 
the correct shape of the cut-off, which is not a 
simple exponential, as assumed by several authors, but rather goes as
$\exp{[-(p/p_{e,\max})^2]}$ \citep{blasi10}. 
This trend is in fact crucial when studying the X-ray emission in the keV band.

Finally, the constant $K_{ep}$ accounts for the different normalization between
electron and proton spectra, very likely related to the different mechanisms
responsible for lepton and hadron injection.
However, since electron injection in SNRs is still far from being understood
from first principles \cite[though see e.g.][]{amano-hoshino, mario-anatoly11} we 
do not fix $K_{ep}$ {\it a priori} but we leave it as a free parameter which
has to be tuned in order to fit the observations, and in particular the
synchrotron emission.

Another important piece of information needed to describe the non-thermal 
emission due to relativistic electrons is their downstream evolution.
After being accelerated at the shock, electrons are in fact advected downstream,
losing energy because of the remnant adiabatic expansion and of radiative
losses (mostly synchrotron, but also ICS). The standard equation describing the
energy evolution from $t_{0}$ to $t$ in this case reads: 
\begin{equation}
 \frac{{\rm d} E}{{\rm d} t} = -\frac{4}{3} \sigma_T c \left( \frac{E}{m_e
   c^2}\right)^2 \frac{B_{\rm eff}^2}{8\pi}
   - \frac{E}{L}\frac{{\rm d}L}{{\rm d}t},
\end{equation}
where $L=L(t_{0},t)$ accounts for adiabatic losses, as in Eq.~(\ref{eq:L}), and
$B_{\rm eff}^2= B^2 + B_{\rm eq}^2$, with $B_{\rm eq} \simeq 12\mu$G is the
equivalent magnetic field which takes into account the IC losses due to CMB radiation,
Galactic background light and IR emission due to local dust (see
\S\ref{sec:gamma}).

Electrons produced at time $t_0$ with energy $E_0$ will thus have an energy 
$E(t)$ at a later time $t$ given by \cite[]{Reynolds98}: 
\begin{equation} \label{eq:Ee_evol}
 \frac{E(t)}{E_{0}} = \frac{L(t_{0},t)}{1 + AE_0 \int_{t_0}^t L(t_{0},\tau)
                      B_{\rm eff}^2(\tau) \, {\rm d}\tau} \,,
\end{equation}
where $A=1.57\times 10^{-3}$ in cgs units.

The electron spectrum at time $t$ can be therefore computed by using number
conservation, namely: $f_e(E) d E= f_{e,0}(E_0) d E_0$. The spectra of
accelerated protons and electrons integrated over the downstream volume, at the
present age of the remnant, are finally showed in Fig.~\ref{fig:p-e}
and in particular the electron spectrum can be described as follows.

For $p\lesssim p_{roll}\simeq 200$ GeV/c, where  $p_{roll}$ is the roll-over
momentum, i.e.\ the momentum for which the radiative loss time becomes
comparable with the SNR age, synchrotron losses are negligible and so electron
and proton spectra are parallel and $\propto p^{-4.2}$.
Between $p_{roll}$ and $p_{cut}\simeq 20$ TeV/c, i.e.\ the momentum for which 
the loss rate exceeds the acceleration one, the electron spectrum is instead
dominated by radiative losses, and therefore goes as $\sim p^{-5.2}$.

It is interesting to notice that the steepening due to synchrotron losses spans
more than one order of magnitude in momentum below $p_{cut}$, therefore it is
expected to have peculiar observational counterparts both in the synchrotron and
in the ICS emissions. We will discuss this topic in details in
\S\ref{sec:roll-off} and \S\ref{sec:gamma}, respectively.

\subsection{Radiative processes}
In order to fit the emission observed from Tycho, we consider the following
processes: 
1) synchrotron emission of relativistic electrons; 
2) thermal and non-thermal electron bremsstrahlung; 
3) ICS of electrons on different microwave, IR and optical photons; 
4) photons due to the decay of neutral pions produced in hadronic collisions. 

The synchrotron emission produced by accelerated electrons in the local
amplified magnetic field (Eq.~\ref{eq:B_damp}) is carried out by using the exact
synchrotron kernel \citep{Rybicki85}.

The bremsstrahlung emission consists of a component due to thermal electrons,
whose temperature is calculated as in \S\ref{sec:electron},
plus a non-thermal component produced by the accelerated ones. 
The total spectrum is the sum of both electron-nucleus and electron-electron 
bremsstrahlung: latter process is however important at the highest energies,
being negligible at lower ones.
For the $e-e$ and the $e-n$ bremsstrahlung we adopt the differential
cross sections provided by \cite{Haug98} and by \cite{Haug97}, respectively.

Electrons contribute to the gamma-ray emission through ICS on local photons.
We consider here the CMB radiation, the IR dust emission and
the Galactic IR + optical background as a function of the distance from the 
Galactic center \citep{Porter05}. 
The flux and spectrum of ICS photons is calculated by using the
exact kernel for ICS, with also the full Klein-Nishina regime accounted for.

Finally, flux and spectrum of the gamma rays produced by $\pi^0$ decay are calculated
following the analytical approximations by \cite{Kelner06}.

\section{Modelling Tycho} \label{sec:results}
In this section and in the next we apply the whole apparatus outlined above to
explain Tycho's properties, along with its multi-wavelength spectrum and
morphology as well.

Our model actually has very few free parameters other than the well-constrained 
ones related with a type Ia SN explosion: the number density of the upstream
medium, $n_0=\rho_{0}/m_{p}$, the injection efficiency $\xi_{\rm inj}$ 
and the electron to proton normalization, $K_{ep}$. The values of $n_0$ and
$\xi_{\rm inj}$ are simultaneously chosen in order to fit the gamma-ray emission
detected by Fermi-LAT and VERITAS, in such a way that $n_{0}$ also allows us to
reproduce the inferred position and velocity of the SNR forward shock.
Eventually, we may regard also the choice of $\chi_{esc}=0.1$ and the assumption
of a constant magnetic field equal to $B_{1}$ upstream as a heuristic way to
account for the maximum hadron energy and, in turn, for the highest-energy
photons detected by VERITAS (see the discussion at the end of
\S\ref{sec:field}).

The best fitting is obtained by adopting $n_0=0.3\, \rm cm^{-3}$ and $\xi_{\rm inj}=3.7$. 
The chosen value for the upstream density is compatible with the upper
limits existing in the literature: \cite{Gamil07} derived $n_0< 0.6\, \rm cm^{-3}$ from the absence
of thermal emission from the shocked ejecta in the {\it Chandra} X-ray data,
while a similar result, $n_0< 0.4\, \rm cm^{-3}$, was obtained by \cite{Volk08}
using the upper limits on the gamma-ray emission measured by HEGRA and Whipple.

The value of $n_0$ can be checked indirectly also by comparing the current
estimates of the remnant's distance and of the shock velocity with
the same quantities computed according to the evolution model explained in
\S~\ref{sec:model}. 
Being Tycho the remnant produced by a type Ia SN, we can safely fix the
explosion energy in $E_{SN}= 10^{51}$ erg and the ejecta mass in about one solar mass.
Therefore, once also the density of the interstellar
medium is fixed, all other quantities related to the SNR evolution are
unequivocally determined.

Our estimate for the distance can be worked out by comparing the final radius of the FS
with the observed size of the remnant: using $n_0=0.3\,\rm cm^{-3}$ we get $d=
3.3$ kpc. 
In the literature different techniques have been adopted to provide an estimate of 
such a distance, always returning a value ranging between 2 and 5 kpc 
\citep[see][for a review]{Hayato10}. Kinematic methods give a
distance of $2.5-3$ kpc \cite[]{Tian11}. It is worth noticing that kinematic
estimates are affected by the fact that Tycho is located in the Perseus arm of
the outer Galaxy, which is influenced by the spiral shock which causes a
velocity reversal. Other methods tends to prefer higher distances. In particular
using the detection of echo light, \cite{Krause08} determine a distance of
$3.8^{+1.5}_{-0.9}$ kpc. Combining the observed ejecta velocities with the
ejecta proper-motion measurements by {\it Chandra}, \cite{Hayato10} derived a
distance of $4\pm 1$ kpc. Finally \cite{Volk08} suggested a lower limit of 3.3
kpc by modeling the existing gamma-ray measurements. In summary our result of
3.3 kpc seem to be consistent with all the existing estimates.

The shock speed that we found at the current age of the remnant is $V_{sh}=
4990$ km s$^{-1}$.  Measuring the proper motion of X-ray rim observed with the
{\it ROSAT} satellite, \cite{Hughes00} derived $V_{sh}= 4600\pm 400 (d/2.3\, \rm
kpc)$ km s$^{-1}$, in perfect agreement with our findings.  Similar results
have been obtained measuring the proper motion in the radio frequencies
\cite[]{Reynoso97}. Another independent technique sometimes used to infer the
shock speed is the measurement of the optical H$\alpha$ emission. 
In fact the line width of the so called {\it broad} H$\alpha$ emission is
related to the temperature of the shocked plasma, which is in turn related to
the shock speed \cite[]{chevalier80}. By adopting this technique \cite{Smith91}
derived shock velocities in the range 1500-2800
km s$^{-1}$, about a factor 2-3 lower than our estimate. 
Nevertheless, we have to notice that the shock speed inferred from the H$\alpha$ line
width is systematically lower than the ones provided by other methods 
in basically all the SNRs considered. 
This discrepancy is probably due to the fact that the downstream
temperature is calculated using a regular hydrodynamical shock, without 
accounting for the possible presence of accelerated particles \citep[see e.g.][]{Helder09} 
and/or the dynamical role of the neutral component.

Finally, we fix the electron to proton ratio from the fit of the synchrotron
emission. Very interestingly, a unique value $K_{ep}= 1.6\times 10^{-3}$ allows
us to fit both the X-ray data from {\it Suzaku} and the radio emission from several experiments \citep{ReyEllison92}.
Our value of $K_{ep}$ is only a factor a few smaller than the ratio
measured at Earth in the diffuse spectrum of Galactic CRs, and such a
discrepancy might be accounted for by remembering that the latter value is the
result of all the SNR evolution (in principle $K_{ep}$ is a function of time)
and it is also averaged over the contributions of different kinds of sources
(e.g.\ type I/II SNRs and maybe pulsars as well). 
At the present time there are some hints about how electron injection 
into DSA may occur \citep{amano-hoshino,mario-anatoly11, Morlino09-inj}, 
but a complete theory coupled to NLDSA and therefore to a self-consistent 
calculation of the non-thermal emission from a SNR during its lifetime is still lacking. 
This is actually the very reason why two of our 
free parameters are somewhat related to the need of
modeling the injection of particles into the acceleration mechanism. 

With both model and environmental parameters fixed, we find that the
Tycho's FS is converting into CRs about the 6 per cent of its bulk
pressure (Fig.~\ref{fig:evo}), which in turn implies a downstream magnetic field
$B_{2}\simeq 300\mu$G. Our shock dynamics is only mildly modified by the
presence of accelerated particles: apart from magnetic field amplification, all
of the signature typical of NLDSA, like the formation of an upstream precursor
where the incoming fluid is slowed down (and in turn the concavity in the
spectrum due to the fact that CRs with different momenta probe different regions
of the precursor) are not very marked. 
For instance, the total and the subshock compression ratios turn out to be 
$R_{tot}=1/U_2\simeq 4.2$, and $R_{sub}=U_1/U_2\simeq 3.9$, respectively. 
This result also justifies \emph{a posteriori} the treatment of the 
SNR evolution as governed by the gas pressure only (\S\ref{sec:hydro}) 
rather than by the whole gas + CRs fluid.

Let us now compare the theoretical prediction for the ratio between 
the positions of CD and FS with the value inferred from the X-ray observations 
by \cite{Warren05} which returned $R_{CD}/R_{sh}= 0.93\pm 2\%$. 
In our model we find $R_{CD}/R_{sh}= 0.87$. 
This value has however to be taken with a grain of salt, since
it may be enhanced by Rayleigh-Taylor instabilities at the CD.
For instance, purely gaseous 1D hydrodynamical simulations predict a significantly smaller
value for this ratio, and namely $R_{CD}/R_{sh}= 0.77$ \cite[]{Wang01}. 
In the same work the authors also performed a 2D simulation and showed 
that the Rayleigh-Taylor instability at the CD allows
fingers of ejecta to protrude well beyond the average CD radius,
also inferring the maximum extent of these fingers in about the 87 per cent of the FS radius. 
Following this line of thought, we can estimate in about a 5 per cent the
boost in the CD/FS position ratio due to instabilities operating at the CD.
When we multiply our prediction by 1.05, we get a CD/FS ratio of about 0.91,
consistently with the value estimated by \cite{Warren05}.

Some authors \cite[e.g.][]{edb05} have proposed that young SNRs can accelerate 
CRs efficiently also at the reverse shock (RS). This idea is supported by the
fact that the RS moves with respect to the unshocked ejecta with a velocity
comparable to that of the FS. 
The RS speed can in fact be estimated from the usual analytical 
model by \cite{TMK99}, and for Tycho it turns out to be $\sim 0.86\, V_{sh}$.
On the other hand, an efficient DSA requires also a sizable magnetic field, say of
order of few $\mu$G, to efficiently scatter particles.
The simple dilution by flux-freezing of the typical magnetic field on the
surface of the progenitor would imply $\sim$nG or lower magnetic field, therefore 
a very efficient magnetic field amplification at the RS would be required to 
make DSA work.
On the other hand, \cite{Warren05} showed that the inferred
distance between the CD and RS is compatible with hydrodynamical model
excluding efficient CR acceleration at the RS.

Anyway, in case of non negligible acceleration at the RS, we would expect an
enhancement of the synchrotron emission only from the position of the RS towards
the inside of the remnant. According to the measurement of \cite{Warren05} the
RS is located at $0.7 R_{sh}$. From Fig.~\ref{fig:radio} we do not notice any
clear excess in this region with respect to our prediction. 
We will discuss in \S\ref{sec:radio} how also our analysis is able to 
account for the observed radio profile without requiring acceleration 
of electrons at the RS. 

\section{Multi-wavelength spectrum} \label{sec:spectrum}
\begin{figure*}
\begin{center}
\includegraphics[width=0.9\textwidth]{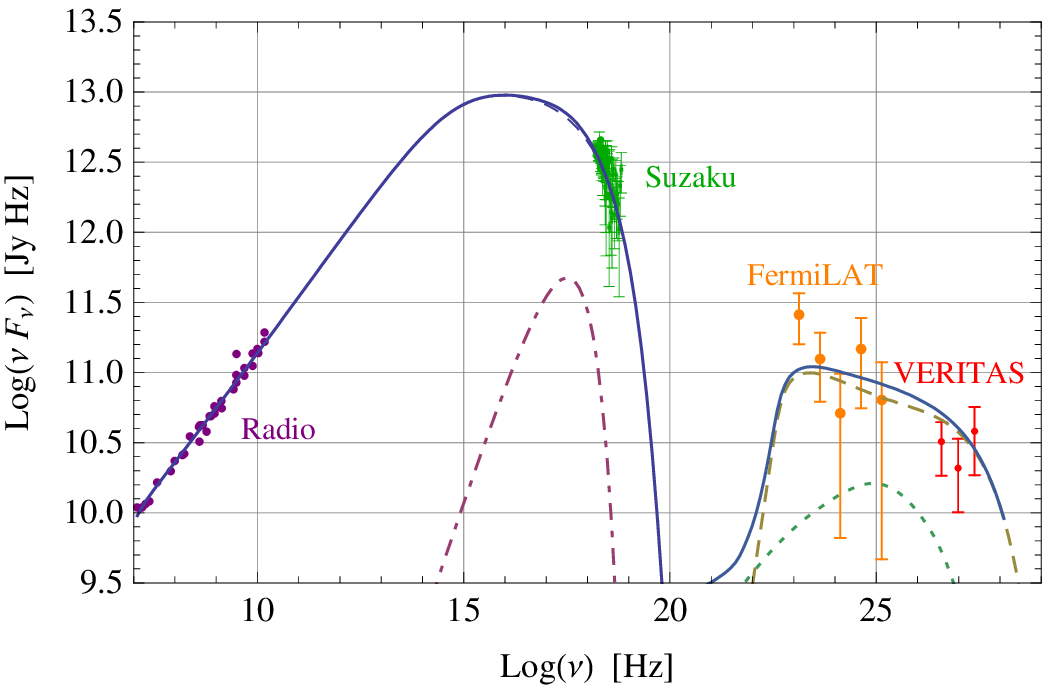}
\caption{Spatially integrated spectral energy distribution of Tycho. 
The curves show synchrotron emission, thermal electron bremsstrahlung and
pion decay as calculated within our model (see text for details). 
The experimental data are, respectivley: radio from {\protect \cite{ReyEllison92}};
X-rays from {\it Suzaku} (courtesy of Toru Tamagawa) , GeV gamma-rays from
Fermi-LAT {\protect \citep{Giordano11}} and TeV gamma-rays from
VERITAS {\protect \citep{Acciari11}}. Both Fermi-LAT and VERITAS data include
only statistical error at 1 $\sigma$. }
\label{fig:multiwave}
\end{center}
\end{figure*}

In Fig.~\ref{fig:multiwave} we show our best fitting of the photon spectrum 
produced by the superposition of all the radiative processes outlined above,
comparing it with the existing data.  The overall agreement is quite good,
therefore we want to analyze now in greater detail the emission in each single
band. 

\subsection{Radio emission} \label{sec:radio}
As depicted in Fig.~\ref{fig:multiwave}, the total radio emission is nicely 
accounted for by our model, which returns a photon spectral index of $0.606$ in
the considered energy range (10 to 1500 MHz).

\begin{figure}
\begin{center}
\includegraphics[width=\linewidth]{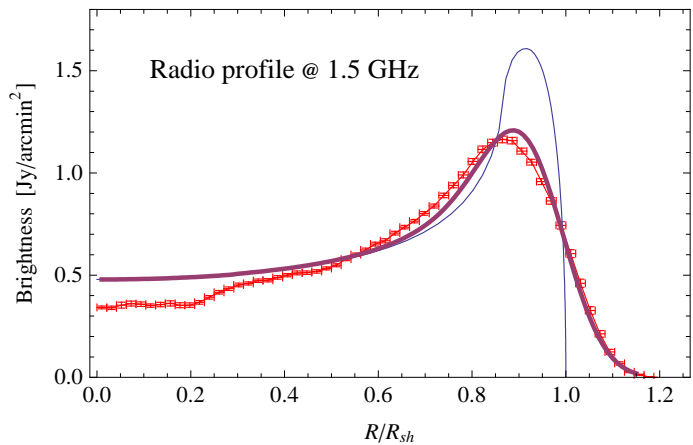}
\caption{Surface brightness of the radio emission at 1.5 GHz as a function of the
radius (data as in Fig.~\ref{fig:Radio_map}). 
The thin solid line represents the projected radial profile computed from our model, while
the thick solid line corresponds to the same profile convoluted with a Gaussian with a
PSF of 15 arcsec.} 
\label{fig:radio}
\end{center}
\end{figure}

Nevertheless, as already noticed by \cite{ReyEllison92}, Tycho's radio emission
shows evidence for a curvature in the spectrum, which turns out to be slightly
steeper (flatter) than $E^{-0.6}$ below (above) 100 MHz. In fact,
\cite{Kothes06} find a spectral index $\alpha = 0.65$ for the radio spectrum
fitted to all flux densities and $\alpha = 0.61$ in the range 408 to 1420 MHz.
\cite{ReyEllison92} ascribed this effect to the curvature of the electron
spectrum due to non-linear effects induced by the CR precursor and stressed the
fact that, in the energy range involved in the radio emission, electron and
proton \emph{energy} spectra are not exactly parallel because protons are
non-relativistic and DSA is \emph{momentum} dependent. 

In our model we did not find any steepening in the lowest energy region of the
electron spectrum, and the reason is the following. The emission in the $1-100$
MHz band, where the curvature is observed, is produced by electrons with energy
close to the injection energy. The typical synchrotron frequency is in fact:
\begin{equation}
 \nu_{\rm syn} \simeq \frac{4 eB \gamma^2}{3 m_e c}= 
     24 \,\frac{B}{100 \mu {\rm G}} \, \frac{\gamma}{10^{2}} \,\rm MHz \,,
\end{equation}
from which, taking the typical injection Lorentz factor $\gamma_{e, {\rm inj}}
\sim 10-100$ and a downstream magnetic field strength of $\sim 100-300 \mu$G,
we obtain $\nu_{\rm syn}\sim 1-100$ MHz. 
The shape of the electron spectrum in this energy range is therefore quite
uncertain, and our simple parametrization of particle injection via thermal
leakage may indeed be too simplistic.
A comparison between thermal leakage model and more refined Monte Carlo
approaches to particle injection, able to retain the smooth transition in the
supra-thermal energy range, seems however to confirm that a proper account for
particle injection should quite naturally predict a spectral steepening in the
10--100 MeV energy range \cite[see fig.~1 of ][]{comparison}.

Therefore, the detection of a curvature in the low-energy radio emission 
from young SNRs \citep{ReyEllison92} may not be necessary related to 
the presence of a CR precursor in the upstream, but rather be a crucial tool 
for probing electron injection. It is also worth stressing that this possibility
is given by the fact that the magnetic field is strongly amplified: for a standard
interstellar field, in fact, the same radio emission would come from
higher-energy electrons, and more precisely in the GeV range.  

Besides the volume-integrated emission, another precious information can be
inferred from the remnant morphology: the radial profile of both radio and X-ray
emissions provides in fact a strong evidence that the magnetic field inside the
SNR is considerably amplified. In order to compare the observed radial profiles,
we need to compute the emission projected along the line of sight. 
As already noticed, Tycho is clearly shell-like both in radio and X-ray bands:
even if the northeastern region shows an enhancement of the emission, 
especially in the radio band, the spherical symmetry can be indeed considered a reasonable
approximation. Under this assumption the local emissivity is a function of the
radius only and the projected emission is thus simply due to the integration
along the line of sight $l$:
\begin{equation} \label{eq:proj}
 j_p(\nu, \rho)= 2
     \int_{\max\left[0,\sqrt{r_{CD}^2-\rho^2}\right]}^{\sqrt{r_{sh} ^2-\rho^2}} 
     j\left(\nu, r=\sqrt{\rho^2+l^2} \right) dl\,.
\end{equation}
the integration limits take into account that the emission is expected to come 
only from the shocked ISM located between the CD and the FS.

In order to be compared with the actual data, the projected emission has to be
convolved with the instrumental point spread function, which we assume to be
Gaussian.
We use the radio map in Fig.~\ref{fig:Radio_map}, which has been obtained with the VLA 
on January 2007 at the frequency of 1.5 GHz, (data from the NRAO/VLA Archive)
and in Fig.~\ref{fig:radio} we compare our results with these data. 
The step
line shows the measured surface brightness obtained by integrating the emission
along the azimuthal angle, between 0 and $2\pi$. The thin solid line shows the
projected emission computed using Eq.~(\ref{eq:proj}), while the thick line is
the emission after the convolution with the instrumental point spread function,
which is equal to 15 arcsec. 

Fig.~\ref{fig:radio} shows indeed a good overall agreement between the data 
and our prediction, even if some little discrepancies can be noticed. 
The most evident one is that in the innermost region ($r/R_{sh}\lesssim 0.3$), 
the theoretical prediction overestimates the observed brightness by about 20 per cent. 
A plausible explanation of this difference may reside in a slight deviation from the 
spherical symmetry, which is somehow expected just because the northeastern region is
brighter than the rest of the remnant. 

Another subtle but interesting difference is that the emission peaks slightly
more inwards than in our model; as a consequence, also the emission detected in
the region $0.6 \lesssim r/R_{sh} \lesssim 0.8$ is found to be a bit larger
than the theoretical prediction. This difference might have different
explanations. The most obvious, and already mentioned, is the possible deviation
from the spherical symmetry. 
Another possibility is given by placing the CD in a different position: 
if one assumed the CD to be located closer to the center
(i.e.\ if one took the CD/FS ratio to be a few per cent smaller),
the theoretical prediction would nicely fit the data. 
However, we can not forget that this explanation would be at odds with
the findings of \cite{Warren05}, who estimated the position of the CD to be
more towards the forward shock, namely around $0.93 R_{sh}$.

A final comment on the radio profile concerns the effects of the non-linear Landau 
damping in the determination of the magnetic field relevant for the synchrotron emission.
If we neglected the damping, the magnetic field strength in the
downstream (dotted line in Fig.~\ref{fig:Bfield}) would lead to a
total radio flux larger by a factor 50 per cent or more with respect to the data, 
even if the radial radio profile would retain a rather similar shape.

\subsection{X-ray emission} \label{sec:X-rays}

\begin{figure}
\begin{center}
\includegraphics[width=\linewidth]{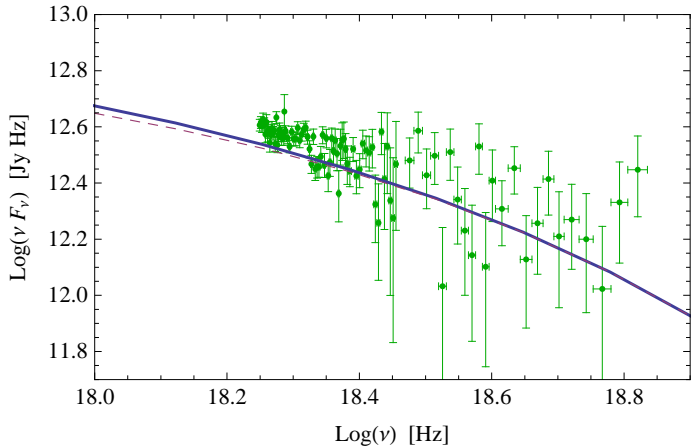}
\caption{X-ray emission due to synchrotron (dashed line) and to synchrotron plus
thermal bremsstrahlung (solid line). Data from the {\it Suzaku} telescope
(courtesy of Toru Tamagawa).} 
\label{fig:X}
\end{center}
\end{figure}

As it is clear from Fig.~\ref{fig:multiwave}, the synchrotron emission spans 
from the radio to the X-ray band, where it sums up with the emission due to 
thermal bremsstrahlung.

The best-fitting to the X-ray continuum observed by {\it Suzaku} data is
illustrated in greater detail in Fig.~\ref{fig:X}, where the dashed line
indicates the synchrotron emission alone and the solid line corresponds to the
sum of synchrotron plus thermal bremsstrahlung.

The electron temperature in the downstream, calculated taking into account
only the heating due to Coulomb collisions with protons (Fig.~\ref{fig:hydrofin}), 
results in a bremsstrahlung emission peaked around 1.2 keV which, at
its maximum, contributes for about the 6 per cent of the total X-ray
continuum emission only, in agreement with the findings of \cite{Gamil07}.

In the same energy range there is however a non-negligible contribution from 
several emission lines, which becomes more and more important moving inwards from 
the FS, where the X-ray emission is mainly non-thermal \citep{Warren05}. 
A detailed model of the line forest is, however,
beyond the main goal of this paper.

\begin{figure}
\begin{center}
\includegraphics[width=\linewidth]{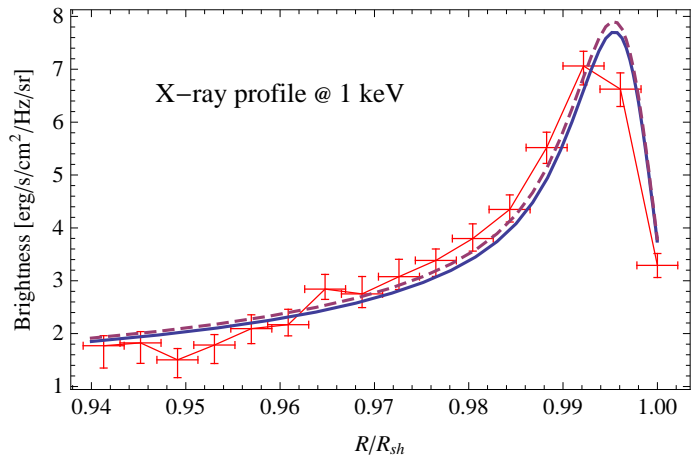}
\caption{Projected X-ray emission at 1 keV. The {\it Chandra} data points are
from {\protect \cite{Gamil07}} (see their Fig.~15). The solid line shows the
projected radial profile of synchrotron emission convolved with the Chandra
point spread function (assumed to be 0.5 arcsec).} 
\label{fig:rim}
\end{center}
\end{figure}

The projected X-ray emission profile, computed at 1 keV, is shown in
Fig.~\ref{fig:rim}, where it is compared with the {\it Chandra} data in the 
region that \cite{Gamil07} call \emph{region W}. The solid curve represents the
resulting radial profile, already convoluted with the {\it Chandra} PSF of about 0.5
arcsec, and shows a remarkable agreement with the data. 
As widely stated above, the sharp decrease of the emission behind the FS is due
to the rapid synchrotron losses of the electrons in a magnetic field as large as
$\sim 300 \mu$G.  In Fig.~\ref{fig:rim} we also plot the radial radio profile
computed without magnetic damping (dashed line); since the typical damping
length-scale is $\sim 3$ pc, it is clear that the non-linear Landau damping can
not contribute to the determination of the filament thickness. 

It is worth stressing that the actual amplitude of the magnetic field we adopt
is not determined to fit the X-ray rim profile, but it is rather a secondary
output, due to our modelling of the streaming instability, of our
tuning the injection efficiency and the ISM density in order to fit the observed gamma-ray
emission (see the discussion in \S\ref{sec:results}). 
We in fact checked {\it a posteriori} whether the corresponding profile of 
the synchrotron emission (which, in shape, is also independent on $K_{ep}$),
were able to account for the thickness of the X-ray rims and for the radio profile as well.

\subsection{Radio to X-ray fitting as a hint of magnetic field amplification}
\label{sec:roll-off}
Another very interesting property of the synchrotron emission is that
a simultaneous fit of both radio and X-ray data may provide a downstream magnetic field 
estimate independent of the one deduced by the rims' thickness. 

In fact, assuming Bohm diffusion, 
the position of the cut-off frequency observed in the X-ray
band turns out to be independent of the magnetic field strength, actually depending
on the shock velocity only. 

On the other hand, if the magnetic field is large enough to make synchrotron
losses dominate on ICS and adiabatic ones, the total X-ray flux in the cut-off
region depends only on the electron density, in turn fixing the value of $K_{ep}$
independently of the magnetic field strength. 
Moreover, radio data suggest the slope of the electron spectrum to be equal to
2.2 at low energies, namely below $E_{roll}\simeq 200$ GeV. Above this energy
the spectral slope has in fact to be 3.2 up to the cut-off determined by setting
the acceleration time equal to the loss time, as discussed in
\S\ref{sec:electron}.

\begin{figure}
\begin{center}
\includegraphics[width=\linewidth]{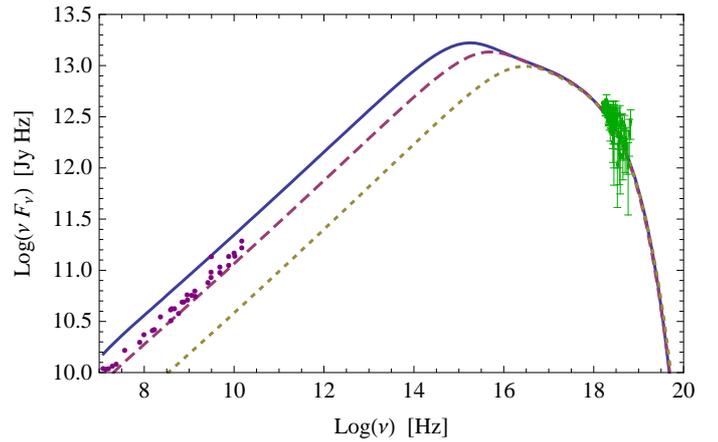}
\caption{Synchrotron emission calculated by assuming constant downstream magnetic
field equal to 100 (dotted line), 200 (dashed line) and 300 $\mu$G (solid line).
The normalization of the electron spectrum is taken to be $K_{ep}=1.6
\times10^{-3}$ for all the curves.} 
\label{fig:sync}
\end{center}
\end{figure}

In Fig.~\ref{fig:sync} we plot the synchrotron emission from the downstream, 
assuming a given magnetic field at the shock and neglecting all the effects 
induced by damping and adiabatic expansion. The three curves correspond to
different values of $B_2=100,200$ and 300$\mu$G, while the normalization factor
$K_{ep}$ is chosen by fitting the X-ray cut-off and it is therefore the same for
all curves. As it is clear from the figure, in order to fit the radio data the
magnetic field at the shock has to be $\gtrsim 200 \mu$G, even in the most
optimistic hypothesis of absence of any damping mechanism acting in the
downstream.

As a matter of fact, synchrotron emission alone can provide an evidence of
ongoing magnetic field amplification, independently of any other evidence
related to X-ray rims' thickness or emission variability. 
Such an analysis is in principle viable for any SNR detected in the non-thermal
X-rays
for which it is also possible to infer the spectral slope of the electron
spectrum from the radio data, only requiring radio and X-ray emissions to come from
the same volume and therefore from the same population of electrons.

\subsection{Gamma-ray emission}\label{sec:gamma}
The most intriguing aspect of Tycho's broadband spectrum is its gamma-ray
emission, which has been detected before in the TeV band by VERITAS
\citep{Acciari11} and then in the GeV band by Fermi-LAT, too \citep{Giordano11}.

Gamma-ray emission from SNRs has been considered for long time a possible
evidence of hadron acceleration in this class of objects \citep{dav94}, even if
there are two distinct physical mechanisms which may be responsible for such an
emission: in the so-called \emph{hadronic scenario} the gamma-rays are produced
by the decay of neutral pions produced in nuclear collisions between CRs and the
background gas, while in the so-called \emph{leptonic scenario} the emission is
due to ICS or relativistic bremsstrahlung of relativistic electrons.

\begin{figure*}
\begin{center}
\includegraphics[width=\textwidth]{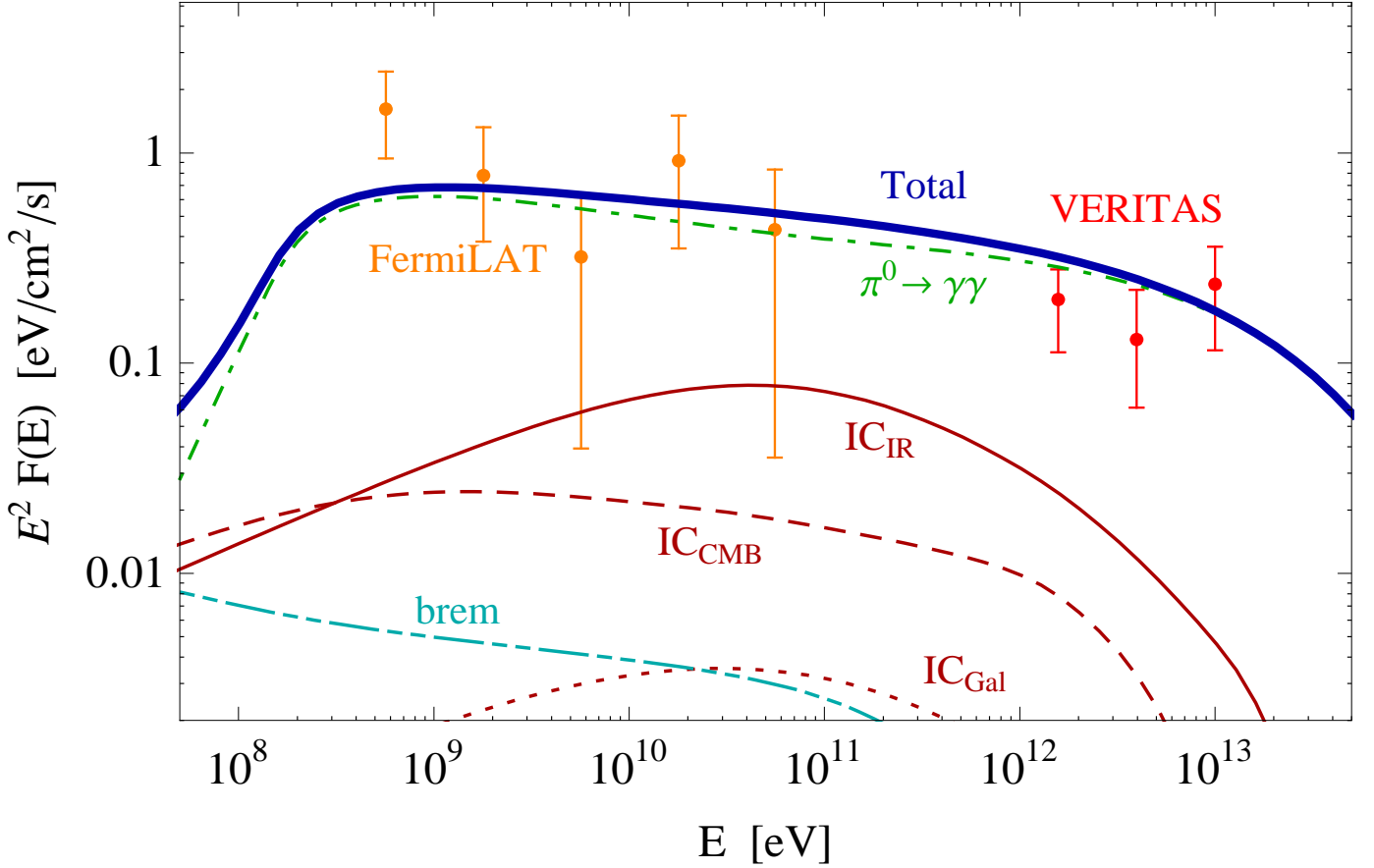}
\caption{Gamma-ray emission observed by Fermi-LAT and by VERITAS compared with 
spectral energy distribution produced by pion decay (dot-dashed line),
relativistic bremsstrahlung (dot-dot-dashed) and ICS computed for three
different photon fields: CMB (dashed), Galactic background (dotted) and IR
photons produced by local warm dust (solid). The thick solid line is the sum of
all the contributions. Both Fermi-LAT and VERITAS data points include only
statistical errors at 1$\sigma$. For VERITAS data the systematic error is found
to be $\sim 30\%$ {\protect {\citep{Acciari11}}}, while for Fermi-LAT the
systematic uncertainties are comparable or even larger than the statistical
error especially for the lowest energy bins due to difficulties in evaluating
the galactic background {\protect\cite[see Fig.~3 in][and the related
discussion]{Giordano11}}.} 
\label{fig:gamma}
\end{center}
\end{figure*}

We show here, with unprecedented clarity for a SNR, that the gamma-ray emission
detected from \T cannot be of leptonic origin, but has to be due to accelerated
hadrons, instead. This fact, along with the VERITAS detection of $\sim10$ TeV
photons and the lack of evidence of a cut-off in the spectrum, implies that
hadrons have to be accelerated up to energies as high as a few hundreds TeV.

In particular, the proton spectrum we obtain shows a cut-off around $p_{\max} =
470$ TeV/c (see Fig.~\ref{fig:p-e}). In this respect, Tycho could be considered
as a \emph{half-PeVatron} at least,  being there no evidence of a cut-off in
VERITAS data. The age-old problem of detecting SNRs emitting photons with
energies larger than a few hundreds TeV (i.e., responsible for the acceleration
of particles up to the knee observed in the spectrum of diffuse Galactic CRs)
may therefore be only a matter of time. Of course it may also be that actually
not all the SNRs are PeVatrons, or even that it is quite unlikely to observe a
SNR in the brief stage of its life in which it may be acting like a PeVatron
\citep{escape,crspec}.

This very important result relies on the fact that the spectral slope in the 
sub-GeV to multi-TeV range is consistent, within measurement errors, with a single
power-law $\propto E^{-2.1}-E^{-2.2}$. Such a power-law is steeper than
the test-particle prediction of DSA and, more importantly, is significantly
different from what expected in the framework of standard NLDSA theories, which
predict concave spectra flatter than $E^{-2}$ at the highest energies \cite[see
e.g.][for some reviews]{drury83,be87,je91,maldrury}.  
   
A key role in our calculations is in fact played by having included in the NLDSA 
also magnetic field amplification, and in particular by having assumed the
relevant Alfv\'en velocity as the one calculated in the amplified magnetic
field. The fact that a small Alfv\'enic Mach number may imply quite steep
spectra had already been put forward by \cite{bell78}, but only by coupling
this fact with a very efficient streaming instability allows NLDSA theories to
predict a proton spectrum steeper than $E^{-2}$ over the whole energy range
\cite[]{zp08,jumpkin}, with also the possibility of changing the spectral slope
by tuning the injection efficiency \cite[]{Texas10}.

In this respect, fixing a value for $\xi_{\rm inj}$ regulates both the slope and
the normalization of the proton spectrum and, in turn, of the gamma-ray
emission. In calculating the photon spectrum produced by $\pi^{0}$ decay
(Fig.~\ref{fig:gamma}) we also accounted for the presence of He nuclei in the
target medium and for the likely presence of accelerated particles heavier than
protons, which are expected to boost the total emission by a factor $\sim 1.4$
and $\sim 1.5$, respectively \citep{Mori09,nuclei}.  

There are two main reasons why the observed gamma-rays cannot be of leptonic
origin: 1) the intensities of both ICS and relativistic bremsstrahlung emission
are too low and 2) the expected shape of the leptonic emission is radically
different from the one observed.

The former point is illustrated in Fig.~\ref{fig:gamma}, where we compare gamma
emissions due to bremsstrahlung, ICS and pion decay. 
The contribution due to relativistic bremsstrahlung (dot-dot-dashed line), however,
is negligible at all the wave-lengths of interest. 
The ICS has been computed for three different target photon families: CMB radiation,
Galactic background radiation and IR due to local dust emission.
The ICS on the CMB radiation (dashed line) provides a contribution a
factor 20--30 smaller than the one by pion decay (dot-dashed line), while the ICS
on the IR + optical Galactic background (dotted line) at Tycho's position
is even smaller. For the Galactic background we used the estimate provided by
\cite{Porter05} for a distance from the Galactic Center of 12 kpc and inside the
Galactic plane, which is a good approximation for Tycho's position 
($\sim$10 kpc away from the Galactic center).

For the ICS we have also included, for the first time as far as we know,
a third contribution due to local dust emission (solid thin line). 
In fact IR radiation from Tycho's direction has been detected by
different satellites in the wavelength band ranging from 10 $\mu$m to 160 $\mu$m
\cite[]{Ishihara10}. 
This emission seems to be due to two distinct populations
of dust, one which we refer to as \emph{warm}, with a temperature $T\sim 100$ K 
and one we refer to as \emph{cold}, with $T\sim 20$K. 
According to \cite{Ishihara10}, the warmer component should originate from
ISM dust collisionally heated by the hotter plasma downstream of the shock. 
In fact, the emission in the mid-IR band shows a shell-like structure 
recalling both synchrotron and H$\alpha$ emissions.
On the other hand, the far-IR emission (140-160 $\mu$m) is dominated 
by a rather cold dust emission correlating with the
position of the CO cloud, which has been showed not to be in physical contact with the SNR itself. 
Hence we consider only the photon background due to the warm dust.
In order to compute the local energy density, we assume that IR photons are
emitted uniformly from the shocked ISM between the CD and the FS. Following
\cite{Ishihara10} we fix the temperature of warm dust to be 100 K, hence, by fitting
the observed IR data, we get a local energy density of $\varepsilon_{\rm warm}= 3.1$
eV/cm$^3$. 
It is important to notice that the ICS on this mid-IR radiation dominates over all
other photon backgrounds. Furthermore we notice that if one assumed also
the far-IR emission (due to cold dust) to originate from a region close to the
remnant, the ICS on these latter photons would be negligible with respect to the 
ICS associated with warm dust photons.
Summing up all the possible photon backgrounds, the resulting total ICS emission
contributes not more than $\sim 20$ per cent of the total gamma emission 
around 0.1 TeV, but it is negligible for lower and higher energies,
as clearly shown in Fig.~\ref{fig:gamma}.

The possibility of explaining the gamma emission using only the ICS on the CMB
photons by assuming a larger value of $K_{ep}$ is inconsistent with the
very efficient magnetic field amplification inferred. 
A $K_{ep}$ smaller by a factor 20-30 would in fact require a downstream magnetic 
field a factor 5-6 smaller in order to keep the synchrotron emission fixed, 
in turn raising issues not only with the radial profile of the synchrotron emission 
(see also \S\ref{sec:X-rays}), but also with the relevant effects induced by a large 
Alfv\'en velocity discussed in \S\ref{sec:field}. 

The second solid argument against a leptonic origin for the detected gamma-ray 
emission is given by the actual slope of the photon spectrum. Once excluded the
underdominant relativistic bremsstrahlung and ICS on the CMB radiation and
Galactic background, we are left with ICS on the IR background due to local dust
as the only viable candidate. 
However, as predicted by standard ICS theory and as showed in 
Fig.~\ref{fig:gamma}, the expected photon spectrum below the
cut-off is typically flatter than parent electrons' one, and more precisely is
$\propto \nu^{-1.6}$ for an electron spectrum $\propto E^{-2.2}$, clearly at
odds with Fermi-LAT data in the GeV range.

Another point worth noticing is that the ICS on the CMB radiation is sensitive
to the steepening of the total electron spectrum above $\sim$100 GeV
(Fig.~\ref{fig:p-e}) due to the synchrotron losses particles undergo while being
advected downstream, while for the ICS on the  IR+optical background the onset
of the Klein-Nishina regime (above $E_{e}\approx 7$ TeV for photons of 1 eV)
does not allow us to probe significantly the steep region of the electron
spectrum.

In other words, ICS on the CMB radiation is too low and cannot be boosted by 
invoking a larger electron density, while ICS on IR and/or optical background, 
which might as well be locally enhanced with respect to the mean Galactic value,
cannot provide a spectral slope in agreement with both Fermi-LAT and VERITAS
data.

We are therefore forced to conclude that the present multi-wavelength analysis
of Tycho's emission represents the best evidence of the fact that SNRs do
accelerate protons, at least up to energies of about 500 TeV. The proton
acceleration efficiency is found to be $\sim 0.06\rho_{0}V_{sh}^{2}$, 
corresponding to converting in CRs a fraction of about 12 per cent of the
kinetic energy density $\frac{1}{2}\rho_{0}V_{sh}^{3}$. As estimated for
instance in \S3 of the review by \cite{Hillas05}, such a value is consistent
with the hypothesis that SNRs are the sources of Galactic CRs, provided that
the residence time in the Milky Way scales with $\sim E^{-1/3}$.

It is important to remember that the actual CRs produced by a single SNR is given
by the convolution over time of different contributions with non trivial spectra,
and namely the flux of particle escaping the remnant from upstream during the 
Sedov-Taylor stages and the bulk of particles released in the ISM at the SNR's death
\citep{escape, crspec}.
In this respect, the instantaneous spectrum of accelerated particles
in Tycho, which is inferred to be as steep as $\propto E^{-2.2}$, provides a hint of
the fact that SNRs can indeed produce rather steep CR spectra as required to
account for the $\propto E^{-2.7}$ diffuse spectrum of Galactic CRs \citep{Texas10}.      

\section{Discussion and conclusions}\label{sec:conclusions}
Thanks to the large amount of data available at different wavelengths,
Tycho's SNR is one of the best object where studying the connection between
CRs and SNRs. Using the observed non-thermal spectrum, and in particular the recent
detection of GeV emission by Fermi-LAT together with the TeV spectrum detected
by VERITAS, we can infer that at Tycho's FS protons are accelerated up to energies
as large as $\sim 500$ TeV, and that the total energy converted into CRs 
can be estimated to be about 12 per cent of the FS bulk kinetic energy.

To reach this conclusion we investigated particle acceleration at the
forward shock using a state-of-the-art semi-analytical NLDSA model including the dynamical
reaction of the accelerated particles, the generation of magnetic fields as due
to streaming instability excited by CRs, the dynamical reaction of these
self-generated fields on the plasma and also the modification of the speed of
the scattering centers (Alfv\'enic drift) induced by the magnetic field amplification
\citep{jumpkin,crspec}. 
The last effect is of crucial importance because it produces a softening of the particle
spectrum with respect to the standard prediction $\propto E^{-2}$,
which allow us to fit both the radio and the gamma photon spectra. 
The stationary version of NLDSA we used is coupled to the dynamical
evolution of the remnant according to the analytical prescription by \cite{TMK99}. 
This allow us to account for information sensitive to the time evolution, 
like for instance the radio to X-ray connection, as illustrated in \S\ref{sec:roll-off}.

We would like to stress that our fit to the observed non-thermal spectrum and 
to the SNR morphology depends on a few parameters only. As for modeling particle
acceleration, we are forced to introduce some parameters describing the
non-yet-understood processes related to the microphysics of the magnetic fields
involved, like the injection of hadrons and electrons (two parameters,
$\xi_{inj}$ and $K_{ep}$) and the scattering of highest-energy particles
($\chi_{esc}$, plus a heuristic choice of a constant magnetic field upstream,
$B_{1}$).

Being Tycho the remnant of a type Ia SN explosion, the only free parameter 
related to the SNR environment is actually the density of the circumstellar
medium where the remnant expands (see \S\ref{sec:results}). 
We adopt a number density  of $n_0=0.3$ cm$^{-3}$, which is consistent with all
of the observational constraints coming from thermal and non-thermal emission
and which also provides position and velocity of the FS in decent agreement with
the estimates provided by several authors adopting different techniques.

Our findings show that the decay of neutral pions produced in hadronic collision between
accelerated ions and gas nuclei is the dominant process in the gamma-ray band.
In particular, we predict a slope for accelerated protons $\propto E^{-2.2}$
which well accounts for Fermi-LAT and VERITAS detections within the
experimental errors. The ICS of relativistic electrons cannot explain the
observed gamma-ray emission, as can be seen from Fig.~\ref{fig:gamma} and as
explained in \S\ref{sec:gamma}. 

The very reasons supporting this conclusion are the following. 
First, the strong magnetic field produced by the CR-induced
streaming instability forces the number density of relativistic electrons to be
too small (the electron to proton ratio is $K_{ep}= 1.6\times 10^{-3}$) to explain
the gamma-ray emission as due to ICS on the ambient photons.
A larger $K_{ep}$ would in fact lead to overestimate the synchrotron emission, 
both in the radio and in the X-ray bands.

Second, even if we arbitrarily reduced the magnetic field strength, enhancing
at the same time the electron number density in order to fit the TeV gamma-rays
with ICS emission, we could not account for the GeV gamma-rays because both the
spectral slope and the flux would be incompatible with the recent Fermi-LAT
observations. 

Also the other competing leptonic process which may show the correct spectral slope 
throughout the whole gamma-ray spectrum, namely the non-thermal bremsstrahlung, 
has to be ruled out, in that it provides a flux two order of magnitudes lower than the Fermi-LAT
detection, and cannot be arbitrarily enhanced without overpredicting both 
the TeV and the X-ray emission.  

In this work we also showed, for the first time as far as we know, that the main contribution to
the ICS is due to the IR photons produced by circumstellar dust heated up through collisions with the
shocked plasma, which dominates on the CMB and Galactic photon background. 
This contribution, which is generally neglected in multi-wavelength studies, may be relevant also for other SNRs,
especially those produced by type Ia SNe which expand in the cold ISM.

As we outlined above, the impossibility of fitting the gamma-ray spectrum 
with leptonic processes is tightly connected with the fact that many observational 
evidences point towards a very efficient magnetic field amplification,
as we comment in the following.

In our model CR-induced streaming instability amplifies the magnetic field
upstream of the shock from the unperturbed Galactic value of $\sim 5\mu$G up to
$\sim 90 \mu$G, which becomes $\sim 300\mu$G immediately downstream because of 
compression at the shock. 
Observationally speaking, such a large magnetic field has two main consequences: 
1) it produces narrow X-ray filaments because electrons undergo severe synchrotron losses; 
2) it determines the roll-over frequency of the synchrotron spectrum to be $\sim 4$ eV/$\hbar$ 
(see \S\ref{sec:roll-off}). 
The former effect has been extensively investigated in the literature and 
it is indeed recovered in our study of the X-ray emission profile (see Fig.~\ref{fig:rim}).

Also the latter effect represents an independent evidence that the magnetic field
has to be amplified up to a few hundreds of $\mu$G.  
As we showed in \S~\ref{sec:roll-off}, when the radio data are good enough to
infer the electron spectral index, the value of $\nu_{roll}$, i.e.\
the frequency emitted by electrons whose loss time equals the age of the remnant, is
uniquely determined by a simultaneous fit of both radio and X-ray data as due to 
synchrotron emission. 
In fact, when synchrotron losses are dominant, $\nu_{roll}$ is a function of the 
average magnetic field downstream $\langle B \rangle$ only, and from
Fig.~\ref{fig:sync} we can see that for Tycho we have $\nu_{roll}\sim 4$
eV/$\hbar$, which in turn implies $\langle B \rangle\simeq 200 \mu$G. 

The narrowness of X-ray filaments might be produced also
by rapid damping of magnetic field downstream of the shock
rather than by synchrotron losses \citep[see e.g.][]{Pohl05}. 
In the case of Tycho we can exclude this possibility simply by
looking at the synchrotron emission: 
if the field were damped below 200 $\mu$G on a length-scale comparable to
the X-ray rim width, we would in fact fail to fit the combination of radio plus
X-ray emission, and the radio emission profile as well. 
This conclusion of course does not imply any damping not to be at work in the 
downstream, but only suggests that no damping mechanisms can determine the structure
of the observed X-ray filaments.

Similar structures in Tycho's interior have been recently detected also 
in the shape of radial stripes by \cite{Eriksen11}. 
These stripes show a surprisingly ordered pattern and have been interpreted 
by the authors as evidences of magnetic fields of few hundreds $\mu$G, compatible with 
acceleration of particles up to the knee.
Nevertheless, we would like to stress that a large magnetic field is necessary but not sufficient
to claim protons to actually be accelerated up to consistently large energies:
we showed NLDSA at the FS to be able to account for protons of about 500 TeV,
i.e. with an energy only a factor a few lower than the knee's in the CR
spectrum.
Larger energies might in principle be achieved at earlier or later times
in Tycho, or even in peculiar region of the FS with a favorable magnetic field
topology.
In any case, we cannot exclude either the possibility that type Ia SNRs, like Tycho, might not be 
responsible for the knee observed in the diffuse spectrum of Galactic CRs, or the possibility that 
we are somehow underestimating $E_{max}$, since VERITAS data do not provide evidence for a high-energy cut-off. 

Some authors have proposed that CR acceleration may occur also at the RS
of young SNRs \citep[see e.g.][]{edb05,ZirAha10,zp11}, even if it is unlikely to
expect a very large magnetic field in the ejecta RS propagate into. 
In the present work we therefore investigated the radial profile of the radio
emission, concluding that there are no evidences of acceleration at the RS. 
In \S\ref{sec:radio} we showed that it is possible to account for the observed radio 
profile including electrons accelerated at the FS only.

In this work we tried to use basically all of the available observations,
interpreting them in the light of a semi-analytical approach to NLDSA to build up
a consistent model of particle acceleration at Tycho's FS. 
However, we are aware that we neglected the information coming from the detection of Balmer lines,
which also point towards efficient particle acceleration \citep{Lee10}. 
The reason is that, at the moment, a reliable NLDSA theory in presence of
neutrals still lacks. 
When shocks propagate into a partially ionized circumstellar
medium, as it might be the case for Tycho, the neutral fraction cannot be 
neglected because it is coupled to the ions via charge-exchange and ionization processes. 
As a consequence, shock dynamics and, in turn, particle acceleration are likely 
sensitive to this aspect of the problem. 
The study of these phenomena might indeed provide new insights on how particle acceleration 
works, but, nevertheless, it is unlikely for these effects 
to invalidate the most important findings of the present work, namely
that Tycho is providing us with the first clear-cut example of hadron
acceleration in SNRs, up to $\sim500$GeV at least.
This result, made possible by the present generation of gamma-ray instruments as
Fermi-LAT and VERITAS, is of primary importance in corroborating the SNR
paradigm for the origin of Galactic CRs.

\begin{acknowledgements}
We would like to thank Gamil Cassam-Chena\"i for useful discussions during a
preliminary phase of this work and for providing us with the {\it Chandra} X-ray
data for Tycho's filaments.
In a similar way, we are glad to thank Toru Tamagawa and Aya Bamba for 
providing us with the {\it Suzaku} X-ray spectrum.
We also want to acknowledge Maite Beltran for helping us in handling the radio
data.
Last, but not the least, we cannot but express our gratitude to Pasquale Blasi 
and Elena Amato for their being a constant source of support and scientific inspiration.
The authors are supported through the contract ASI-INAF I/088/06/0 (grant
TH-037).
\end{acknowledgements}

\end{document}